\documentclass[10pt,journal,final]{IEEEtran}
\IEEEoverridecommandlockouts

\usepackage{amsfonts,amssymb}
\usepackage{ifpdf}
\usepackage{cite}

\usepackage[hyphens]{url}
\usepackage{stfloats}
\usepackage{bm}
\usepackage{color,xcolor}
\usepackage{subfigure}
\usepackage{caption}

\ifCLASSINFOpdf
   \usepackage[pdftex]{graphicx}
   \graphicspath{{../pdf/}{../jpeg/}}
   \DeclareGraphicsExtensions{.pdf,.jpeg,.png}
\else
   \usepackage[dvips]{graphicx}

   \graphicspath{{../eps/}}

   \DeclareGraphicsExtensions{.eps}
\fi

\usepackage[cmex10]{amsmath}

\usepackage{algorithm}
\usepackage{algorithmic}
\ifCLASSOPTIONcompsoc
\else
  \usepackage[caption=false,font=footnotesize]{subfig}
\fi
\usepackage{makecell}
\begin{document}
%\title{User-Adaptive IRS Beamforming Design  for IoT Applications with Diverse Requirements}
%\title{A Simple Architecture  Design of Semi-Passive Intelligent Reflecting Surface  for Localization}
\title{Intelligent Reflecting Surface Assisted   Localization: Performance Analysis and Algorithm Design}
\author{Meng Hua, ~\IEEEmembership{Member,~IEEE,}
%	\thanks{This work was supported by the Education Planning Project of Hunan Province under Grant ND212050, the General Project of Natural Science Foundation of Hunan Province (Research on Energy Efficiency Theory and Key Technology of Data Collection System Based on Fixed-wing UAV), the Open Project of Key Laboratory of Intelligent Control Technology for Wuling-Mountain Ecological Agriculture in Hunan Province under Grant ZNKPY2019-1, the Key Project of Department of Education of Hunan Province under Grant 20A394, the National Natural Science Foundation of China under Grant 62061017, the Department of Education of Hunan Province under Grant 19B456.  The associate editor coordinating the review of this paper and approving it for publication was Dr. Qingqing Wu. (\emph{Corresponding author: Qingheng~Song}.)}
%Luxi~Yang,~\IEEEmembership{Senior Member,~IEEE,}
%Qingqing~Wu,
%Cunhua~Pan,
Qingqing~Wu,~\IEEEmembership{Senior Member,~IEEE,}
 Wen~Chen,~\IEEEmembership{Senior Member,~IEEE,}
Zesong~Fei,~\IEEEmembership{Senior Member,~IEEE,}
Hing Cheung So,~\IEEEmembership{Fellow,~IEEE,}
and Chau~Yuen,~\IEEEmembership{Fellow,~IEEE}
%%
%%%\thanks{Copyright (c) 2015 IEEE. Personal use of this material is permitted. However, permission to use this material for any other purposes must be obtained from the IEEE by sending a request to pubs-permissions@ieee.org.}
%%\thanks{Manuscript received April   27, 2019; revised July    29, and accepted November  7, 2019. This work was supported by National Natural Science Foundation of China under Grant  61971128,  Grant 61372101, and Grant 61671144, Scientific Research Foundation of Graduate School of Southeast University  under Grand  YBPY1859 and China Scholarship Council (CSC) Scholarship, National High Technology Project of China  under 2015AA01A703,  Cyrus Tang Foundation Endowed Young Scholar Program under SEU-CyrusTang-201801.   The associate editor coordinating the review of this paper and approving it for publication was Kamel Tourki. (\emph{Corresponding author: Luxi Yang}.)}

%\thanks{   L. Yang and C. Li are with the School of Information Science and Engineering, Southeast University, Nanjing 210096, China (e-mail: \{  lxyang, chunguoli\}@seu.edu.cn).}
%\thanks{Q. Wu is with the State Key Laboratory of Internet of Things for Smart City and Department of Electrical and Computer Engineering, University of Macau, Macao 999078 China (email: qingqingwu@um.edu.mo). }
%\thanks{C. Pan is with the School of Electronic Engineering	and Computer Science, Queen Mary University of London, London E1 4NS, U.K. (e-mail: c.pan@qmul.ac.uk).}
\thanks{M. Hua is with the Department of Electronic Engineering, Shanghai Jiao Tong University, 200240, China, and also with the State Key Laboratory of Internet of Things for Smart City, University of Macau, Macau, 999078, China (email: menghua@um.edu.mo).}

\thanks{Q. Wu and W. Chen are with the Department of
	Electronic Engineering, Shanghai Jiao Tong University, 200240, China (email: \{qingqingwu,wenchen\}@sjtu.edu.cn). }

\thanks{Z. Fei is with the School of Information and
	Electronics, Beijing Institute of Technology, Beijing 100081, China (email: feizesong@bit.edu.cn). }

\thanks{HC. So is with the Department of Electrical Engineering, City University of Hong Kong, Hong Kong, 999077, China (e-mail: hcso@ee.cityu.edu.hk).}
\thanks{C. Yuen is with the School of
	Electrical and Electronics Engineering, Nanyang Technological University,  639798, Singapore  (e-mail: chau.yuen@ntu.edu.sg).}
}
\vspace{-1.5cm}
%\vspace{-2truemm}
\maketitle
\begin{abstract}
The target sensing/localization performance   is fundamentally limited by  the line-of-sight link and severe signal  attenuation over long distances. This paper considers a challenging scenario where the direct link between the base station (BS) and the target is blocked due to the surrounding blockages and leverages the  intelligent reflecting surface (IRS) with some active sensors, termed as \textit{semi-passive IRS}, for localization. To be specific, the active sensors receive echo signals reflected by the target and apply signal processing techniques to estimate the target location.  We consider the joint time-of-arrival (ToA) and direction-of-arrival (DoA) estimation  for localization  and derive the corresponding  Cram\'{e}r-Rao bound (CRB), and then a simple  ToA/DoA estimator without iteration is proposed. In particular, the  relationships of the CRB for ToA/DoA with the number of frames for IRS beam adjustments,   number of IRS reflecting elements, and   number of sensors are theoretically analyzed and demystified. Simulation results show that the proposed semi-passive IRS architecture   provides sub-meter level positioning accuracy even over a long localization range from the BS to the target and also demonstrate  a  significant localization accuracy improvement compared to the fully passive IRS architecture. 
\end{abstract}
\begin{IEEEkeywords}
Intelligent reflecting surface (IRS),  target  localization, time-of-arrival (ToA), direction-of-arrival (DoA), Cram\'{e}r-Rao bound (CRB).
\end{IEEEkeywords}

\section{Introduction}
%With the emerging environment-aware applications such as smart transportation, autonomous driving, pedestrian/animal intrusion detection, and  unmanned aerial vehicle   trajectory tracking, 
%the  research of enabling high-resolution and high-accuracy sensing is  attracting  great attention and has been  initialized by  the 3rd Generation Partnership Project (3GPP) \cite{onlineweb}. In   traditional sensing/localization, the base station (BS) transmits the probing signal and then receives the echo signal reflected by the targets for estimation \cite{so2011source}. However, it is difficult to achieve  high  sensing performance  in complex  ground-based environments due to the following reasons. First, the line-of-sight (LoS)  of the BS-target link  is usually absent in urban areas  due to   surrounding obstacles such as buildings and transportation vehicles. Second, the received echo power at the BS is weak due to the limited BS transmit power and severe signal attenuation over long distances. 

With the emerging environment-aware applications such as smart transportation, 
the  research of enabling high-resolution and high-accuracy sensing is  attracting  great attention and has been  initialized by  the 3rd Generation Partnership Project (3GPP) \cite{onlineweb}. In   traditional sensing/localization, the base station (BS) transmits the probing signal and then receives the echo signal reflected by the targets for estimation \cite{so2011source}. However, it is difficult to achieve  high  sensing performance  in complex  ground-based environments due to the absence of line-of-sight (LoS) channel and severe signal attenuation over long distances.

Recently,  the emerging technology, namely, intelligent reflecting surface (IRS),  has been envisioned as  an effective way to address the above two issues. The IRS  is composed of a large number of passive reflecting elements, which has the ability to  change the wireless propagation environment \cite{qingqingwu2021intelligent,chen2022active,zhou2020Intelligent,9913311},
%. Due to this appealing property, it has been reported in \cite{chen2022active,zhou2020Intelligent,hua2021Intelligent}
% that  a significant throughput can be achieved  when  integrating IRS into current networks.  In fact, 
 the IRS is also beneficial for  sensing \cite{meng2022Intelligent,shao2022target,Keykhosravi2021SISO,Elzanaty2021Reconfigurable,hu2023IRS}.  Due to its easy deployment, the IRS can be attached to the facades of buildings or streetlights so that the virtual LoS link can be established between the BS and  target. For example,  \cite{Keykhosravi2021SISO}  proposed to leverage the \textit{fully-passive IRS} for localization where the BS  receives signals reflected by IRS for direction-of-arrival (DoA) and time-of-arrival (ToA) estimation. It  showed that  a sub-meter level positioning is  achieved. However, this accuracy is attained only for  short sensing ranges since the probing signal travels  through the BS-IRS-target-IRS-BS link incurring very high signal attenuation. To overcome this issue,   \cite{shao2022target}
proposed a \textit{semi-passive IRS} architecture where some active sensors are integrated into the IRS so that the signal traveling through the BS-IRS-target-sensor link can be processed at the sensor side, which substantially reduces the signal attenuation. It was shown  that accurate estimation of the  target direction can be achieved even in   large sensing ranges. However,   \cite{shao2022target} only focuses on DoA  estimation, the target  location cannot be obtained. Therefore, the joint DoA and ToA estimation is required for localizing target. However, it imposes two new challenges. First, the ToA estimation significantly differs from the DoA estimation, which indicates that the new algorithm is required for ToA estimation under the semi-passive IRS architecture. Second, the performance limit of  the semi-passive IRS architecture for localization is unknown. Specifically, the impact of system parameters including  the number of frames for IRS beam adjustments,   number of IRS reflecting elements, and   number of sensors on localization performance remains unknown.
% but does not estimate the ToA, leading to two research questions to be addressed. First, the impact of system parameters including  the number of frames for IRS beam adjustments,   number of IRS reflecting elements, and   number of sensors on localization performance remains unknown.  Second, joint DoA and ToA estimation has not been investigated under the semi-passive IRS architecture.
%On the other hand, how to design  efficient estimators  to jointly estimate  DoA and  ToA under the semi-passive architecture is also unknown. 
\begin{figure}[!t]
	\centerline{\includegraphics[width=2.5in]{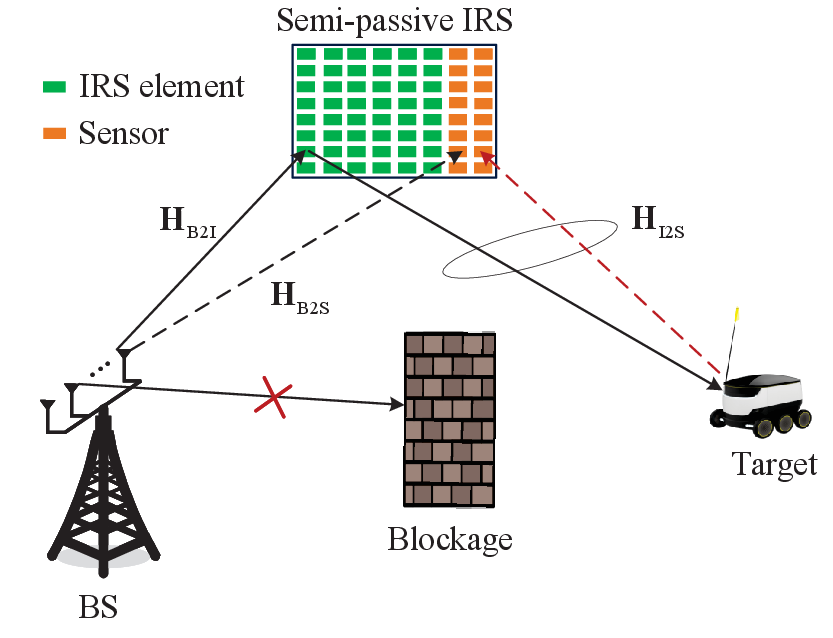}}
	\caption{Semi-passive IRS aided localization system.} \label{fig1}
	\vspace{-0.5cm}
\end{figure}

In this paper, we tackle the above two challenges by applying  the semi-passive IRS architecture for localization.  To this end, we first derive Cram\'{e}r-Rao bound (CRB) to
characterize a lower bound of the mean squared error (MSE) of the target location and then unveil the relationship between CRB and system parameters.  Particularly, the optimal numbers of IRS reflecting elements and sensors  to minimize the ToA/DoA estimation error are derived.
Then, we   propose efficient estimators to estimate the target location. To be specific, the MUSIC algorithm is applied to  estimate the target DoA. Then, by transforming the received sensor data into   frequency-domain sequences  via discrete Fourier transform (DFT), we apply the maximum likelihood (ML) estimator   to jointly estimate the ToA and the  target-related coefficient. The advantage is that  these two steps do not require any iterations, thereby resulting in  low computation.
Simulation results show that   sub-meter level positioning accuracy is achieved  even over a long distance from the BS to the target. Moreover,  the superiority of  semi-passive IRS architecture over the fully-passive IRS  is also verified.
\vspace{-0.2cm}
\section{System Model}
We consider a localization system consisting of one BS, a semi-passive IRS, and one target with  unknown location as shown in Fig.~\ref{fig1}. The direct link between the BS and   target is blocked due to the surrounding obstacles  around the BS.\footnote{Our proposed design is also applicable to the case with the direct link between the BS and the target with some modifications  on distinguishing multi-path signals. In this work, we  focus on the non-line-of-sight case only in order to obtain useful insights.}  
%\footnote{ \color{red}For the case where the direct link between the BS and the target exists,  the analysis for target localization  will become more challenging since the multiple path signals should be distinguished  at the sensors. In this work, we  focus on the non-line-of-sight case only in order to obtain useful insights here and leave the challenging case as our future work. } 
The coordinates of the BS, the semi-passive IRS, and the target are denoted by 
${{\bf{q}}_{{\rm{BS}}}} = {\left[ {{x_{{\rm{BS}}}},{y_{{\rm{BS}}}},{z_{{\rm{BS}}}}} \right]^T}$, ${{\bf{q}}_{\rm{I}}} = {\left[ {{x_{\rm{I}}},{y_{\rm{I}}},{z_{\rm{I}}}} \right]^T}$, and  ${{\bf{q}}_{\rm{u}}} = {\left[ {{x_{\rm{u}}},{y_{\rm{u}}},0} \right]^T}$, respectively.
The semi-passive IRS consists of two parts: reflecting elements and active sensors, where  reflecting elements are passive and only used for reconfiguring the phases of incident signals,  while the sensors are active and have  signal processing abilities. The BS  has ${{N_{{\rm{BS}}}}}$ antennas along the $y$ axis and the semi-passive IRS has $N_{\rm r}$ reflecting elements and $N_{\rm s}$ sensors  lying on the $y\text{-}o\text{-}z$ plane.\footnote{
	Although we consider  a uniform linear array at the IRS for ease of exposition, our proposed algorithm is also applicable to the case with a uniform planar array since the horizontal and vertical directions can be separately estimated \cite{shao2022target}.}  We consider a  quasi-static flat-fading channel in which the channel state information  remains unchanged in the considered frames.
%\begin{figure}[!t]
%	\centerline{\includegraphics[width=2.8in]{IRSadjustment.eps}}
%	\caption{Illustration of  semi-passive IRS aided sensing system.} \label{fig2}
%%	\vspace{-0.5cm}
%\end{figure}
\vspace{-0.3cm}
\subsection{Channel Model}
As shown in Fig.~\ref{fig1}, the sensors receive signals from two links: BS-IRS (passive side)-target-sensor link and BS-sensor link. Their  channel models are described as follows.

The BS-IRS (passive side) channel is modeled as 
\begin{align}
{{\bf{H}}_{{\rm{B2I}}}} = {\beta _{{\rm{B2I}}}}{{\bf{b}}_{\rm{r}}}\left( {\mu _{{\rm{B2I}}}^{\rm{A}}} \right){{\bf{a}}^H}\left( {\mu _{{\rm{B2I}}}^{\rm{D}}} \right),
\end{align}
where ${\beta _{{\rm{B2I}}}}{\rm{ = }}\sqrt {\frac{{{\lambda ^2}}}{{16{\pi ^2}d_{{\rm{B2I}}}^2}}} $ denotes the large-scale path loss, $\lambda $ represents  carrier wavelength, ${d_{{\rm{B2I}}}}$ stands for the distance between the BS and the IRS, ${\mu _{{\rm{B2I}}}^{\rm{D}}}$ and ${\mu _{{\rm{B2I}}}^{\rm{A}}}$ denote the frequency azimuth angle-of-departure (AoD) from the BS to the IRS and the frequency azimuth angle-of-arrival (AoA) at the IRS, respectively, 
and ${\bf{a}}\left( {\mu _{{\rm{B2I}}}^{\rm{D}}} \right)$ and ${{\bf{b}}_{\rm{r}}}\left( {\mu _{{\rm{B2I}}}^{\rm{A}}} \right)$ denote the  transmit array response vector of the BS and the receive array response vector of IRS given by ${\bf{a}}\left( {\mu _{{\rm{B2I}}}^{\rm{D}}} \right){\rm{ = }}{\left[ {{e^{ - \frac{{j\left( {{N_{{\rm{BS}}}} - 1} \right)\pi \mu _{{\rm{B2I}}}^{\rm{D}}}}{2}}}, \ldots ,{e^{\frac{{j\left( {{N_{{\rm{BS}}}} - 1} \right)\pi \mu _{{\rm{B2I}}}^{\rm{D}}}}{2}}}} \right]^T}$ and  ${{\bf{b}}_{\rm{r}}}\left( {\mu _{{\rm{B2I}}}^{\rm{A}}} \right){\rm{ = }}{\left[ {{e^{ - \frac{{j\left( {{N_{\rm{r}}} - 1} \right)\pi \mu _{{\rm{B2I}}}^{\rm{A}}}}{2}}}, \ldots ,{e^{\frac{{j\left( {{N_{\rm{r}}} - 1} \right)\pi \mu _{{\rm{B2I}}}^{\rm{A}}}}{2}}}} \right]^T}$, respectively. The channel  between the BS and the sensors, i.e., ${{{\bf{H}}_{{\rm{B2S}}}}}$,  can be modeled similarly, and thus it is omitted here.

The IRS-target-sensor channel is modeled as 
\begin{align}
{{\bf{H}}_{{\rm{I2S}}}} = \alpha {\beta _{{\rm{I2S}}}}{{\bf{b}}_{\rm{s}}}\left( {\mu _{{\rm{I2U}}}^{\rm{A}}} \right){\bf{b}}_{\rm{r}}^T\left( {\mu _{{\rm{I2U}}}^{\rm{D}}} \right),
\end{align}
where ${\beta _{{\rm{I2S}}}}{\rm{ = }}\sqrt {\frac{{{\lambda ^2}{\kappa _{{\rm{RCS}}}}}}{{64{\pi ^3}d_{{\rm{I2U}}}^4}}} $, ${d_{{\rm{I2U}}}}$ denotes the distance between the IRS and the target, ${{\kappa _{{\rm{RCS}}}}}$ represents the radar cross-section, $\alpha\sim {\cal CN}\left( {0,1} \right)$ denotes the small fading caused by the fluctuation of the target. In addition, 
${{\bf{b}}_{\rm{s}}}\left( {\mu _{{\rm{I2U}}}^{\rm{A}}} \right){\rm{ = }}{\left[ {{e^{ - \frac{{j\left( {{N_{\rm{s}}} - 1} \right)\pi \mu _{{\rm{I2U}}}^{\rm{A}}}}{2}}}, \ldots ,{e^{\frac{{j\left( {{N_{\rm{s}}} - 1} \right)\pi \mu _{{\rm{I2U}}}^{\rm{A}}}}{2}}}} \right]^T}$ and ${{\bf{b}}_{\rm{r}}}\left( {\mu _{{\rm{I2U}}}^{\rm{D}}} \right){\rm{ = }}{\left[ {{e^{ - \frac{{j\left( {{N_{\rm{r}}} - 1} \right)\pi \mu _{{\rm{I2U}}}^{\rm{D}}}}{2}}}, \ldots ,{e^{\frac{{j\left( {{N_{\rm{r}}} - 1} \right)\pi \mu _{{\rm{I2U}}}^{\rm{D}}}}{2}}}} \right]^T}$. We consider that the target is in the far field of the semi-passive IRS, the AoD from the IRS (passive side) to the target equals the AoA from the target to the sensors. For notational simplicity,  we denote $\mu _{{\rm{I2U}}}$ by 
$\mu _{{\rm{I2U}}}{\rm{ = }}\mu _{{\rm{I2U}}}^{\rm{A}}{\rm{ = }}\mu _{{\rm{I2U}}}^{\rm{D}}$. 
\vspace{-0.3cm}
\subsection{IRS-aided Sensing Protocol}
Since  the target position is unknown to the BS, the IRS should adjust different beams  with different directions to scan the target. 
%  The proposed protocol of IRS for sensing target is shown in Fig.~\ref{fig2}, where the time channel coherence is divided into
   Without loss of generality, we assume that there are  $N_{\rm f}$ frames and in each frame, the IRS adjusts beam once. Let  ${\bf{\Theta }}\left( n \right){\rm{ = diag}}\left( {{\bm{\theta }}\left( n \right)} \right)$ with  ${\bm{\theta }}\left( n \right) = {\left[ {{e^{j{\theta _1}\left( n \right)}}, \ldots ,{e^{j{\theta _{{N_{\rm{r}}}}}\left( n \right)}}} \right]^T}$ denote the IRS phase shift matrix in the $n$th frame.

The  signal  received by the sensor at time   $t$ in the  $n$th frame is given by 
\begin{align}
{\bf{\tilde y}}\left( {n,t} \right)& = {{\bf{H}}_{{\rm{I2S}}}}{\bf{\Theta }}\left( n \right){{\bf{H}}_{{\rm{B2I}}}}{\bf{w}}s\left( t-\tau_{\rm tot} \right) \notag\\
&+{{\bf{H}}_{{\rm{B2S}}}}{\bf{w}}s\left( {t - {\tau _{{\rm{B2I}}}}} \right) + {\bf{ n}}\left( t \right), \forall n,  \label{received_signa}
\end{align}
where  $s\left( t \right)$ stands for the transmitted signal, ${\bf{w}}$ represents the BS beamformer,  ${\tau _{{\rm{tot}}}} = {\tau _{{\rm{B2I}}}} + 2{\tau _{{\rm{I2U}}}}$, ${{\tau _{{\rm{B2I}}}}}$ and ${{\tau _{{\rm{I2U}}}}}$ denote ToAs from the BS to  the IRS and from the IRS to the target, respectively,  and ${\bf{ n}}\left( t \right)$ is the  circularly symmetric complex Gaussian noise satisfying ${\bf{n}}\left( t \right) \sim {\cal CN}\left( {0,{\sigma ^2}{{\bf{I}}_{{N_{\rm{s}}}}}} \right)$.  As the locations of the  BS and the semi-passive IRS are known, the channel information ${{\bf{H}}_{{\rm{B2S}}}}{\bf{w}}s\left( {t - {\tau _{{\rm{B2I}}}}} \right)$  can be   known by the sensor in advance via offline estimation  and can be perfectly canceled before performing target location estimation at the sensor side.\footnote{Note that the signal from the BS to the sensor does not contain any useful information about the target, which indicates that no ``Fisher information'' is provided by the BS-sensor link and is removed here to improve estimation accuracy.} 

Thus, after  removing ${{\bf{H}}_{{\rm{B2S}}}}{\bf{w}}s\left( {t - {\tau _{{\rm{B2I}}}}} \right)$ and performing  algebraic operations in \eqref{received_signa},   we can transform \eqref{received_signa} into 
\begin{align}
 {\bf{y}}\left( {n,t} \right) & = \alpha {\beta _{{\rm{B2I}}}}{\beta _{{\rm{I2S}}}}{{\bf{b}}_{\rm{s}}}\left( {\mu _{{\rm{I2U}}}^{}} \right){\bf{b}}_{\rm{r}}^T\left( {\mu _{{\rm{I2U}}}^{}} \right){\bf{\Theta }}\left( n \right){{\bf{b}}_{\rm{r}}}\left( {\mu _{{\rm{B2I}}}^{\rm{A}}} \right)\notag\\
&\times{{\bf{a}}^H}\left( {\mu _{{\rm{B2I}}}^{\rm{D}}} \right){\bf{w}}s\left( {t - {\tau _{{\rm{tot}}}}} \right) + {\bf{n}}\left( t \right), \forall n. \label{received_signa_1}
\end{align}
 Obviously, to maximize the power received at the sensor, the BS should directly adjust the beam towards the IRS, i.e., the optimal beamformer $\bf w$ is set as 
\begin{align}
{\bf{w}}^{\rm opt} = \sqrt {\frac{{{P_{{\rm{BS}}}}}}{{{N_{{\rm{BS}}}}}}} {\bf{a}}\left( {\mu _{{\rm{B2I}}}^{\rm{D}}} \right), \label{optimal_beamformer}
\end{align}
where ${{P_{{\rm{BS}}}}}$ denotes the BS transmit power.

Then, substituting \eqref{optimal_beamformer} into  \eqref{received_signa_1}  yields  
 \begin{align}
 {\bf{y}}\left( {n,t} \right)&={\beta _{{\rm{target}}}}{{\bf{b}}_{\rm{s}}}\left( {\mu _{{\rm{I2U}}}^{}} \right){\bf{b}}_{\rm{r}}^T\left( {\mu _{{\rm{I2U}}}^{}} \right){\bf{\Theta }}\left( n \right){{\bf{b}}_{\rm{r}}}\left( {\mu _{{\rm{B2I}}}^{\rm{A}}} \right)\notag\\
 &\times s\left( {t - {\tau _{{\rm{tot}}}}} \right) + {\bf{n}}\left( t \right), \label{received_signa_2}
 \end{align}
where ${\beta _{{\rm{target}}}} = \alpha \sqrt {{P_{{\rm{BS}}}}{N_{{\rm{BS}}}}} {\beta _{{\rm{B2I}}}}{\beta _{{\rm{I2S}}}}$.
\vspace{-0.3cm}
\subsection{Location Estimation Performance Evaluation via CRB}
Since the target location estimation error is difficult to quantify in general, we introduce the CRB as the performance metric, which is essentially  a lower bound of any unbiased estimator.

We define a vector of unknown parameters ${{\bf{u}}_{{\rm{position}}}}{\rm{ = }}{\left[ {{x_{\rm{u}}},y_{\rm{u}}^{},\beta _{{\rm{target}}}^{{\mathop{\rm Re}\nolimits} },\beta _{{\rm{target}}}^{{\mathop{\rm Im}\nolimits} }} \right]^T}$, where ${\beta _{{\rm{target}}}^{{\mathop{\rm Re}\nolimits} }}$ and 
${\beta _{{\rm{target}}}^{{\mathop{\rm Im}\nolimits} }}$ denote the real and imaginary  parts  of ${\beta _{{\rm{target}}}}$, respectively. However, it is difficult to derive the Fisher information matrix (FIM) of ${{\bf{u}}_{{\rm{position}}}}$ (denoted by ${{\bf{F}}_{{\rm{position}}}}$), instead we first compute the FIM of vector ${{\bf{u}}_{{\rm{channel}}}}{\rm{ = }}{\left[ {{\tau _{{\rm{tot}}}},\mu _{{\rm{I2U}}}^{},\beta _{{\rm{target}}}^{{\mathop{\rm Re}\nolimits} },\beta _{{\rm{target}}}^{{\mathop{\rm Im}\nolimits} }} \right]^T}$ (denoted by ${{\bf{F}}_{{\rm{channel}}}}$), and then use the chain rule to derive the FIM of ${{\bf{u}}_{{\rm{position}}}}$ as follows:
\begin{align}
{{\bf{F}}_{{\rm{position}}}}=\frac{{\partial {{\bf{u}}_{{\rm{channel}}}}}}{{\partial {{\bf{u}}_{{\rm{position}}}}}}{{\bf{F}}_{{\rm{channel}}}}{\left( {\frac{{\partial {{\bf{u}}_{{\rm{channel}}}}}}{{\partial {{\bf{u}}_{{\rm{position}}}}}}} \right)^T}, \label{fishermatrix_position}
\end{align}
where $\frac{{\partial {{\bf{u}}_{{\rm{channel}}}}}}{{\partial {{\bf{u}}_{{\rm{position}}}}}}{\rm{ = }}\left[ {\frac{{\partial {\tau _{{\rm{tot}}}}}}{{\partial {{\bf{u}}_{{\rm{position}}}}}}\frac{{\partial \mu _{{\rm{I2U}}}^{}}}{{\partial {{\bf{u}}_{{\rm{position}}}}}}\frac{{\partial \beta _{{\rm{target}}}^{{\mathop{\rm Re}\nolimits} }}}{{\partial {{\bf{u}}_{{\rm{position}}}}}},\frac{{\partial \beta _{{\rm{target}}}^{{\mathop{\rm Im}\nolimits} }}}{{\partial {{\bf{u}}_{{\rm{position}}}}}}} \right]$. Since there are $N_{\rm f}$ frames for localization, we   rewrite ${{\bf{F}}_{{\rm{channel}}}}$ into  a sum of $N_{\rm f}$ independent FIMs as ${{\bf{F}}_{{\rm{channel}}}} = \sum\limits_{n = 1}^{{N_{\rm{f}}}} {{{\bf{F}}_{{\rm{channel}}}}\left( n \right)} $. In the following, we  show how to  calculate ${{{\bf{F}}_{{\rm{channel}}}}\left( n \right)},\forall n$.

The probability density function of the received signal $ {\bf{y}}\left( {n,t} \right)$ in \eqref{received_signa_2} conditioned on ${{{\bf{u}}_{{\rm{channel}}}}}$ is given by \cite{fu2009cramer} 
\begin{align}
%&p\left( {{\bf{y}}\left( {n,t} \right);{{\bf{u}}_{{\rm{channel}}}}} \right) = \notag\\
%&\qquad\qquad\quad~~{\eta }\exp \left( { - \frac{1}{{{n_0}}}\int {{{\left\| {{\bf{y}}\left( {n,t} \right) - {\bf{\bar y}}\left( {n,t} \right)} \right\|}^2}{\rm{d}}t} } \right),
&p\left( {{\bf{y}}\left( {n,t} \right)} \right) = {\eta }\exp \left( { - \frac{1}{{{n_0}}}\int {{{\left\| {{\bf{y}}\left( {n,t} \right) - {\bf{\bar y}}\left( {n,t} \right)} \right\|}^2}{\rm{d}}t} } \right),
\end{align} 
where  ${ n}_0$ denotes the noise power density, $\eta$ is a constant which has no contributions on ${{\bf{u}}_{{\rm{position}}}}$, and ${\bf{\bar y}}\left( {n,t} \right)={\beta _{{\rm{target}}}}{{\bf{b}}_{\rm{s}}}\left( {\mu _{{\rm{I2U}}}^{}} \right){\bf{b}}_{\rm{r}}^T\left( {\mu _{{\rm{I2U}}}^{}} \right){\bf{\Theta }}\left( n \right){{\bf{b}}_{\rm{r}}}\left( {\mu _{{\rm{B2I}}}^{\rm{A}}} \right)s\left( {t - {\tau _{{\rm{tot}}}}} \right)$. Thus,  FIM ${{\bf{F}}_{{\rm{channel}}}}\left( n \right)$ can be computed as 
\begin{align}
{{\bf{F}}_{{\rm{channel}}}}\left( n \right)=\left[ {\begin{array}{*{20}{c}}
	{{F_{\tau \tau }}\left( n \right)}&{{F_{\tau \mu }}\left( n \right)}&{{{\bf{F}}_{\tau {\bm{\beta }}}}\left( n \right)}\\
	{F_{\tau \mu }^T\left( n \right)}&{{F_{\mu \mu }}\left( n \right)}&{{{\bf{F}}_{\mu {\bm{\beta }}}}\left( n \right)}\\
	{F_{\tau {\bm \beta} }^T\left( n \right)}&{F_{\mu {\bm \beta} }^T\left( n \right)}&{{{\bf{F}}_{{\bm{\beta \beta }}}}\left( n \right)}
	\end{array}} \right], \label{Fisher_frame}
\end{align}
where ${\bm{\beta }}={\left[ {\beta _{{\rm{target}}}^{{\rm{Re}}},\beta _{{\rm{target}}}^{{\rm{Im}}}} \right]^T}$. The $(i,j)$   entry of  ${{\bf{F}}_{{\rm{channel}}}}\left( n \right)$ is calculated by  ${{\bf{F}}_{{\rm{channel}}}}{\left( n \right)_{i,j}} =  - {\mathbb E}\left\{ {\frac{{{\partial ^2}\ln p\left( {{\bf{y}}\left( {n,t} \right);{{\bf{u}}_{{\rm{channel}}}}} \right)}}{{\partial {{\bf{u}}_{{\rm{channel,}}i}}\partial {{\bf{u}}_{{\rm{channel,}}j}}}}} \right\}$. 
The following  notation is defined for later use: $\dot s\left( {t - {\tau _{{\rm{tot}}}}} \right){\rm{ = }}\frac{{\partial s\left( {t - {\tau _{{\rm{tot}}}}} \right)}}{{\partial \tau }}$, $\ddot s\left( {t - {\tau _{{\rm{tot}}}}} \right){\rm{ = }}\frac{{{\partial ^2}s\left( {t - {\tau _{{\rm{tot}}}}} \right)}}{{{\partial ^2}\tau }}$, ${{\bf{\dot b}}_{\rm{s}}}\left( {\mu _{{\rm{I2U}}}^{}} \right) = \frac{{\partial {{\bf{b}}_{\rm{s}}}\left( {\mu _{{\rm{I2U}}}^{}} \right)}}{{\mu _{{\rm{I2U}}}^{}}}$, ${{\bf{\ddot b}}_{\rm{s}}}\left( {\mu _{{\rm{I2U}}}^{}} \right) = \frac{{{\partial ^2}{{\bf{b}}_{\rm{s}}}\left( {\mu _{{\rm{I2U}}}^{}} \right)}}{{{\partial ^2}\mu _{{\rm{I2U}}}^{}}}$,  ${\bf{\dot b}}_{\rm{r}}^{}\left( {\mu _{{\rm{I2U}}}^{}} \right) = \frac{{\partial {{\bf{b}}_{\rm{r}}}\left( {\mu _{{\rm{I2U}}}^{}} \right)}}{{\mu _{{\rm{I2U}}}^{}}}$, ${E_{\rm{s}}} = \int {{{\left| {s\left( {n,t - {\tau _{{\rm{tot}}}}} \right)} \right|}^2}{\rm{d}}t} $, and  ${{\bar W}^2} = \int {{{\left| {\frac{{\partial s\left( {n,t - {\tau _{{\rm{tot}}}}} \right)}}{{\partial \tau }}} \right|}^2}{\rm{d}}t/{E_{\rm{s}}}} $. Note that  we have the following identities:  ${\bf{b}}_{\rm{s}}^H\left( {\mu _{{\rm{I2U}}}^{}} \right){{\bf{\ddot b}}_{\rm{s}}}\left( {\mu _{{\rm{I2U}}}^{}} \right) =  - \frac{{{\pi ^2}{N_{\rm{s}}}\left( {{N_{\rm{s}}} - 1} \right)\left( {{N_{\rm{s}}} + 1} \right)}}{{12}}$, ${\bf{b}}_{\rm{r}}^H\left( {\mu _{{\rm{I2U}}}^{}} \right){{{\bf{\dot b}}}_{\rm{r}}}\left( {\mu _{{\rm{I2U}}}^{}} \right) = 0$, and ${\bf{b}}_{\rm{s}}^H\left( {\mu _{{\rm{I2U}}}^{}} \right){{{\bf{\dot b}}}_{\rm{s}}}\left( {\mu _{{\rm{I2U}}}^{}} \right) = 0$.  Then, we can calculate 
${{F_{\tau \tau }}\left( n \right)}$, ${{F_{\tau \mu }}\left( n \right)}$, ${{{\bf{F}}_{\tau \bm \beta }}\left( n \right)}$, ${{F_{\mu \mu }}\left( n \right)}$, ${{{\bf{F}}_{\mu \bm \beta }}\left( n \right)}$, and ${{{\bf{F}}_{\bm \beta \bm \beta }}\left( n \right)}$ in \eqref{Fisher_frame} respectively as 
\begin{align}
&{{{F}}_{\tau \tau }}\left( n \right){\rm{ = }}\frac{{{\rm{2}}{{\left| {{\beta _{{\rm{target}}}}} \right|}^2}{N_{\rm{s}}}}}{{{n_0}}}{\left| {{\bf{b}}_{\rm{r}}^T\left( {\mu _{{\rm{I2U}}}^{}} \right){\bf{\Theta }}\left( n \right){{\bf{b}}_{\rm{r}}}\left( {\mu _{{\rm{B2I}}}^{\rm{A}}} \right)} \right|^2}{{\bar W}^2}{E_{\rm{s}}}, \label{fisher_tau}\\
&{{{F}}_{\tau \mu }}\left( n \right){\rm{ = }}\frac{{{\rm{2}}{{\left| {{\beta _{{\rm{target}}}}} \right|}^2}{N_{\rm{s}}}}}{{{n_0}}}\int {{\mathop{\rm Re}\nolimits} \left\{ {{\bf{\dot b}}_{\rm{r}}^T\left( {\mu _{{\rm{I2U}}}^{}} \right){\bf{\Theta }}\left( n \right){{\bf{b}}_{\rm{r}}}\left( {\mu _{{\rm{B2I}}}^{\rm{A}}} \right)} \right.} \times \notag\\
& \left. {{\bf{b}}_{\rm{r}}^H\left( {\mu _{{\rm{B2I}}}^{\rm{A}}} \right){{\bf{\Theta }}^H}\left( n \right){\bf{b}}_{\rm{r}}^{\rm{*}}\left( {\mu _{{\rm{I2U}}}^{}} \right)s\left( {t - {\tau _{{\rm{tot}}}}} \right){{\dot s}^{\rm{*}}}\left( {t - {\tau _{{\rm{tot}}}}} \right)} \right\}{\rm{d}}t,\\
&{{{F}}_{\tau \beta _{{\rm{target}}}^{{\mathop{\rm Re}\nolimits} }}}\left( n \right){\rm{ = }}\frac{{{\rm{2}}{N_{\rm{s}}}{{\left| {{\bf{b}}_{\rm{r}}^T\left( {\mu _{{\rm{I2U}}}^{}} \right){\bf{\Theta }}\left( n \right){{\bf{b}}_{\rm{r}}}\left( {\mu _{{\rm{B2I}}}^{\rm{A}}} \right)} \right|}^2}}}{{{n_0}}}\notag\\
&\quad \times {\mathop{\rm Re}\nolimits} \left\{ {\beta _{{\rm{target}}}^{}\int {\dot s\left( {n,t - {\tau _{{\rm{tot}}}}} \right){s^{\rm{*}}}\left( {n,t - {\tau _{{\rm{tot}}}}} \right){\rm{d}}t} } \right\},\\
&{{{F}}_{\tau \beta _{{\rm{target}}}^{{\mathop{\rm Im}\nolimits} }}}\left( n \right) = \frac{{{\rm{ - 2}}{N_{\rm{s}}}{{\left| {{\bf{b}}_{\rm{r}}^T\left( {\mu _{{\rm{I2U}}}^{}} \right){\bf{\Theta }}\left( n \right){{\bf{b}}_{\rm{r}}}\left( {\mu _{{\rm{B2I}}}^{\rm{A}}} \right)} \right|}^2}}}{{{n_0}}}\notag\\
&\quad \times {\mathop{\rm Re}\nolimits} \left\{ {j\beta _{{\rm{target}}}^{}\int {\dot s\left( {n,t - {\tau _{{\rm{tot}}}}} \right){s^{\rm{*}}}\left( {n,t - {\tau _{{\rm{tot}}}}} \right){\rm{d}}t} } \right\},\\
&{{{F}}_{\mu \mu }}\left( n \right) = \frac{{{E_{\rm{s}}}{{\left| {{\beta _{{\rm{target}}}}} \right|}^2}{N_{\rm{s}}}}}{{6{n_0}}}\left( {{\rm{1}}2{{\left| {{\bf{\dot b}}_{\rm{r}}^T\left( {\mu _{{\rm{I2U}}}^{}} \right){\bf{\Theta }}\left( n \right){{\bf{b}}_{\rm{r}}}\left( {\mu _{{\rm{B2I}}}^{\rm{A}}} \right)} \right|}^2}} \right.\notag\\
&+ {{\pi ^2}\left( {{N_{\rm{s}}} - 1} \right)\left( {{N_{\rm{s}}} + 1} \right)\left| {{\bf{b}}_{\rm{r}}^T\left( {\mu _{{\rm{I2U}}}^{}} \right){\bf{\Theta }}\left( n \right){{\bf{b}}_{\rm{r}}}\left( {\mu _{{\rm{B2I}}}^{\rm{A}}} \right)} \right|}\Big),\\
&{{{F}}_{\mu \beta _{{\rm{target}}}^{{\mathop{\rm Re}\nolimits} }}}\left( n \right){\rm{ = }}\frac{{2{N_{\rm{s}}}{E_{\rm{s}}}}}{{{n_0}}}{\rm{Re}}\left\{ {\beta _{{\rm{target}}}^{}{\bf{b}}_{\rm{r}}^H\left( {\mu _{{\rm{B2I}}}^{\rm{A}}} \right){{\bf{\Theta }}^H}\left( n \right){\bf{b}}_{\rm{r}}^{\rm{*}}\left( {\mu _{{\rm{I2U}}}^{}} \right)} \right.\notag\\
&\qquad\qquad\quad\times \left. {{\bf{\dot b}}_{\rm{r}}^T\left( {\mu _{{\rm{I2U}}}^{}} \right){\bf{\Theta }}\left( n \right){{\bf{b}}_{\rm{r}}}\left( {\mu _{{\rm{B2I}}}^{\rm{A}}} \right)} \right\},\\
&{{{F}}_{\mu \beta _{{\rm{target}}}^{{\mathop{\rm Im}\nolimits} }}}\left( n \right){\rm{ = }}\frac{{2{N_{\rm{s}}}{E_{\rm{s}}}}}{{{n_0}}}{\rm{Re}}\left\{ {\beta _{{\rm{target}}}^{*}{\bf{b}}_{\rm{r}}^H\left( {\mu _{{\rm{B2I}}}^{\rm{A}}} \right){{\bf{\Theta }}^H}\left( n \right){\bf{b}}_{\rm{r}}^{\rm{*}}\left( {\mu _{{\rm{I2U}}}^{}} \right)} \right.\notag\\
&\qquad\qquad\quad \times \left. {{\bf{\dot b}}_{\rm{r}}^T\left( {\mu _{{\rm{I2U}}}^{}} \right){\bf{\Theta }}\left( n \right){{\bf{b}}_{\rm{r}}}\left( {\mu _{{\rm{B2I}}}^{\rm{A}}} \right)} \right\},
\end{align}
\begin{align}
&{{{F}}_{\beta _{{\rm{target}}}^{{\mathop{\rm Re}\nolimits} }\beta _{{\rm{target}}}^{{\mathop{\rm Re}\nolimits} }}}\left( n \right){\rm{ = }}\frac{{{\rm{2}}{N_{\rm{s}}}{{\left| {{\bf{b}}_{\rm{r}}^T\left( {\mu _{{\rm{I2U}}}^{}} \right){\bf{\Theta }}\left( t \right){{\bf{b}}_{\rm{r}}}\left( {\mu _{{\rm{B2I}}}^{\rm{A}}} \right)} \right|}^2}{E_{\rm{s}}}}}{{{n_0}}},\\
&{{{F}}_{\beta _{{\rm{target}}}^{{\mathop{\rm Re}\nolimits} }\beta _{{\rm{target}}}^{{\mathop{\rm Im}\nolimits} }}}\left( n \right)=0,\\
&{{{F}}_{\beta _{{\rm{target}}}^{{\mathop{\rm Im}\nolimits} }\beta _{{\rm{target}}}^{{\mathop{\rm Im}\nolimits} }}}\left( n \right){\rm{ = }}\frac{{{\rm{2}}{N_{\rm{s}}}{{\left| {{\bf{b}}_{\rm{r}}^T\left( {\mu _{{\rm{I2U}}}^{}} \right){\bf{\Theta }}\left( t \right){{\bf{b}}_{\rm{r}}}\left( {\mu _{{\rm{B2I}}}^{\rm{A}}} \right)} \right|}^2}{E_{\rm{s}}}}}{{{n_0}}}.\label{fisher_beta_imag}
\end{align}
As a result, based on \eqref{fishermatrix_position} and \eqref{fisher_tau}-\eqref{fisher_beta_imag}, the CRB of target position ${\left( {{x_{\rm u}},{y_{\rm u}}} \right)^T}$  is given by 
\begin{align}
{\rm{CRB}}\left[ {{{\left( {{x_{\rm u}},{y_{\rm u}}} \right)}^T}} \right] = \sum\nolimits_{i = 1}^2 {{{\left[ {{\bf{F}}_{{\rm{position}}}^{{\rm{ - }}1}} \right]}_{i,i}}}. \label{CRB_location}
\end{align}
\subsection{Performance Analysis}
 The CRB matrix of  ${{{\bf{u}}_{{\rm{channel}}}}}$ is the inverse of the sum of ${{\bf{F}}_{{\rm{channel}}}}\left( n \right)$ in \eqref{Fisher_frame} given by 
\begin{align}
{\rm{CRB}}\left[ {{{\bf{u}}_{{\rm{channel}}}}} \right]{\rm{ = }}{\left( {\sum\nolimits_{n = 1}^{{N_{\rm{f}}}} {{{\bf{F}}_{{\rm{channel}}}}\left( n \right)} } \right)^{ - 1}}. \label{CRB_channel}
\end{align}
For the special case where the IRS phase shift matrix is set as ${{\bf{\Theta }}^{{\rm{opt}}}}\left( n \right) = \mathop {\arg \max }\limits_{{\bf{\Theta }}\left( n \right)} \left| {{\bf{b}}_{\rm{r}}^T\left( {\mu _{{\rm{I2U}}}^{}} \right){\bf{\Theta }}\left( n \right){{\bf{b}}_{\rm{r}}}\left( {\mu _{{\rm{B2I}}}^{\rm{A}}} \right)} \right|$ to maximize the received signal power at the sensors, we have the  following proposition.

\textbf{\emph{Proposition }1:} With  IRS phase shift matrix  ${{\bf{\Theta }}^{{\rm{opt}}}}\left( n \right)$ and  $\int {\dot s\left( {t - {\tau _{{\rm{tot}}}}} \right){s^{\rm{*}}}\left( {t - {\tau _{{\rm{tot}}}}} \right){\rm{d}}t} {\rm{ = }}0$,
we can obtain the closed-form CRBs  of   $\tau_{\rm tot}$ and  $\mu_{\rm I2U}$ as 
\begin{align}
	&{\rm{CRB}}\left( {{\tau _{{\rm{tot}}}}} \right)=\frac{{{n_0}}}{{{\rm{2}}{{\left| {{\beta _{{\rm{target}}}}} \right|}^2}{N_{\rm{s}}}N_{\rm{r}}^2{N_{\rm{f}}}{{\bar W}^2}{E_{\rm{s}}}}},\label{CRB_tau}\\
	&{\rm{CRB}}\left( {{\mu _{{\rm{I2U}}}}} \right)=\frac{{6{n_0}}}{{{\pi ^2}{N_{\rm{s}}}\left( {{N_{\rm{s}}} - 1} \right)\left( {{N_{\rm{s}}} + 1} \right)N_{\rm{r}}^2{N_{\rm{f}}}{{\left| {{\beta _{{\rm{target}}}}} \right|}^2}{E_{\rm{s}}}}}.\label{CRB_mu}
\end{align}
\emph{Proof:} The  solution of $\mathop {\arg \max }\limits_{{\bf{\Theta }}\left( n \right)} \left| {{\bf{b}}_{\rm{r}}^T\left( {\mu _{{\rm{I2U}}}^{}} \right){\bf{\Theta }}\left( n \right){{\bf{b}}_{\rm{r}}}\left( {\mu _{{\rm{B2I}}}^{\rm{A}}} \right)} \right|$ is given by  ${{\bm{\theta }}^{{\rm{opt}}}}\left( n \right){\rm{ = }}\frac{{{{\left( {{\rm{diag}}\left( {{\bf{b}}_{\rm{r}}^T\left( {\mu _{{\rm{I2U}}}^{}} \right)} \right){{\bf{b}}_{\rm{r}}}\left( {\mu _{{\rm{B2I}}}^{\rm{A}}} \right)} \right)}^*}}}{{\left| {{\rm{diag}}\left( {{\bf{b}}_{\rm{r}}^T\left( {\mu _{{\rm{I2U}}}^{}} \right)} \right){{\bf{b}}_{\rm{r}}}\left( {\mu _{{\rm{B2I}}}^{\rm{A}}} \right)} \right|}}$.  
Then, we  have ${\left| {{\bf{b}}_{\rm{r}}^T\left( {\mu _{{\rm{I2U}}}^{}} \right){{\bf{\Theta }}^{{\rm{opt}}}}\left( n \right){{\bf{b}}_{\rm{r}}}\left( {\mu _{{\rm{B2I}}}^{\rm{A}}} \right)} \right|^2}=N_{\rm{r}}^2$ and ${\bf{\dot b}}_{\rm{r}}^T\left( {\mu _{{\rm{I2U}}}^{}} \right){{\bf{\Theta }}^{{\rm{opt}}}}\left( n \right){{\bf{b}}_{\rm{r}}}\left( {\mu _{{\rm{B2I}}}^{\rm{A}}} \right) = 0$. Together with $\int {\dot s\left( {t - {\tau _{{\rm{tot}}}}} \right){s^{\rm{*}}}\left( {t - {\tau _{{\rm{tot}}}}} \right){\rm{d}}t} {\rm{ = }}0$, we   have  ${{{F}}_{\tau \mu }}\left( n \right)=0$, ${{{F}}_{\tau \beta _{{\rm{target}}}^{{\mathop{\rm Re}\nolimits} }}}\left( n \right)=0$, ${{{F}}_{\tau \beta _{{\rm{target}}}^{{\mathop{\rm Im}\nolimits} }}}\left( n \right)=0$, ${{{F}}_{\mu \beta _{{\rm{target}}}^{{\mathop{\rm Re}\nolimits} }}}\left( n \right)=0$, and ${{{F}}_{\mu \beta _{{\rm{target}}}^{{\mathop{\rm Im}\nolimits} }}}\left( n \right)=0$. As a result, \eqref{CRB_channel} reduces  to  a diagonal matrix, thus leading to the desired result. 

%The results in \eqref{CRB_tau} and \eqref{CRB_mu} unveil  the potential gains brought by the IRS and sensors. To be specific, the CRB of ${{\tau _{{\rm{tot}}}}}$ in \eqref{CRB_tau} shows that the ToA  estimation error is  inversely proportional to ${N_{\rm{s}}}N_{\rm{r}}^2$. T

Observing from \eqref{CRB_tau}, we can see that the CRB of ${{\tau _{{\rm{tot}}}}}$  is  inversely proportional to ${N_{\rm{s}}}N_{\rm{r}}^2$. This  result can be explained as follows. Since the ToA estimation error is only related to the  power received at the  sensors  regardless of the  direction of the  signal impinging on the sensors. Moreover, it can be seen from  \eqref{received_signa_2} that the received power comes from two aspects. The first aspect is the linear receive beamforming gain  of   ${\cal O}\left( {{N_{\rm{s}}}} \right)$ brought by   sensors and the other is the passive beamforming gain of  ${\cal O}\left( {{N_{\rm{r}}^2}} \right)$ brought by the IRS.
Different from ToA estimation, the CRB of ${{\mu _{{\rm{I2U}}}}}$ in \eqref{CRB_tau} shows that the DoA estimation error is  inversely proportional to ${N_{\rm{s}}}\left( {{N_{\rm{s}}} - 1} \right)\left( {{N_{\rm{s}}} + 1} \right)N_{\rm{r}}^2$ since  it is related to both the  received power and  the direction of the  signal arriving at the sensors.
 More specifically,  the  IRS provides the passive beamforming gain of ${\cal O}\left( {{N_{\rm{r}}^2}} \right)$, while the  the sensors not only provide the linear receive beamforming gain but  also the spatial direction gain of ${\cal O}\left( {N_{\rm{s}}^3} \right)$ as in the conventional radar systems \cite{so2011source}.
 
In some applications, we may be interested in  either ToA or DoA. The optimal system parameter configuration for each objective can be obtained. Specifically,  provided that the total number of active sensors and reflecting elements is fixed (denoted by $N$), i.e., ${N_{\rm{s}}} + {N_{\rm{r}}} = {N}$,  the optimal ${N_{\rm{r}}}$ to minimize ${\rm{CRB}}\left( {{\tau _{{\rm{tot}}}}} \right)$ is given by ${2/3N}$. In addition, the optimal ${N_{\rm{r}}}$ to minimize ${\rm{CRB}}\left( {{\mu _{{\rm{I2U}}}}} \right)$ is given by $N_{\rm{r}}^{{\rm{opt}}} = \sqrt[3]{{\left( {{N^3} - 5N} \right)/125 + \sqrt { {N^4} - 2{N^2} - 5} /25}} + \sqrt[3]{{\left( {{N^3} - 5N} \right)/125 - \sqrt {  {N^4} - 2{N^2} - 5} /25}}$. With optimal ${N_{\rm{r}}^{\rm opt}}$, the optimal ${N_{\rm{s}}}$ is given by $N - N_{\rm{r}}^{{\rm{opt}}}$.\footnote{
When considering the practical integer constraint, the optimal solutions  ${N_{\rm{r}}}$ and ${N_{\rm{s}}}$ can be directly obtained from the above solutions.}

\section{Target Location Estimation Algorithm Design}
In this section, we propose a low-complexity algorithm for estimating   DoA ${{\mu _{{\rm{I2U}}}}}$, ToA ${{\tau _{{\rm{tot}}}}}$, and parameter ${\beta _{{\rm{target}}}}$.
\vspace{-0.3cm}
\subsection{Estimate of DoA ${{\mu _{{\rm{I2U}}}}}$}
We first stack the received signals  in \eqref{received_signa_2} as a matrix form given by 
\begin{align}
%&{\bf{Y}}\left( t \right){\rm{ = }}\left[ {{\bf{y}}\left( {1,t} \right),{\bf{y}}\left( {2,t} \right), \ldots ,{\bf{y}}\left( {{N_{\rm{f}}},t} \right)} \right]\notag\\
%&={{\bf{b}}_{\rm{s}}}\left( {\mu _{{\rm{I2U}}}^{}} \right)\left( {{\beta _{{\rm{target}}}}{\bf{b}}_{\rm{r}}^T\left( {\mu _{{\rm{I2U}}}^{}} \right)\left[ {{\bf{\Theta }}\left( 1 \right), \ldots ,{\bf{\Theta }}\left( {{N_{\rm{f}}}} \right)} \right]{{\bf{b}}_{\rm{r}}}\left( {\mu _{{\rm{B2I}}}^{\rm{A}}} \right)} \right)\notag\\
%&\times s\left( {t - {\tau _{{\rm{tot}}}}} \right){\rm{ + }}{\bf{N}}\left( t \right).
{\bf{Y}}\left( t \right)&={{\bf{b}}_{\rm{s}}}\left( {\mu _{{\rm{I2U}}}^{}} \right)\left( {{\beta _{{\rm{target}}}}{\bf{b}}_{\rm{r}}^T\left( {\mu _{{\rm{I2U}}}^{}} \right)\left[ {{\bf{\Theta }}\left( 1 \right), \ldots ,{\bf{\Theta }}\left( {{N_{\rm{f}}}} \right)} \right]} \right)\notag\\
&\times {{\bf{b}}_{\rm{r}}}\left( {\mu _{{\rm{B2I}}}^{\rm{A}}} \right)s\left( {t - {\tau _{{\rm{tot}}}}} \right){\rm{ + }}{\bf{N}}\left( t \right).
\end{align} 
The auto-correlation matrix of ${{\bf{y}}\left( {n,t} \right)}$ is approximately as
\begin{align}
{{\bf{R}}_{{\bf{Y}}\left( t \right)}}{\rm{ = }}\frac{1}{{{N_{\rm{f}}}}}{\bf{Y}}\left( t \right){{\bf{Y}}^H}\left( t \right).
\end{align} 
As there is only one target direction,   the largest eigenvalue of ${{\bf{R}}_{{\bf{Y}}\left( t \right)}}$
spans the signal subspace and the remaining $N_{\rm s}-1$ eigenvectors correspond to  the noise subspace. Specifically, performing singular value decomposition on ${{\bf{R}}_{{\bf{Y}}\left( t \right)}}$, we have 
\begin{align}
%{{\bf{R}}_{{\bf{Y}}\left( t \right)}}{\rm{ = }}\left[ {{{\bf{U}}_{\rm{s}}}{\kern 1pt} {\kern 1pt} {\kern 1pt} {{\bf{U}}_{\rm{n}}}} \right]\left[ {\begin{array}{*{20}{c}}
%	{{{{\Sigma }}_{\rm{s}}}{\kern 1pt} {\kern 1pt} {\kern 1pt} }&{}\\
%	{}&{{{\bf{\Sigma }}_{\rm{n}}}{\kern 1pt} {\kern 1pt} }
%	\end{array}} \right]{\left[ {{{\bf{U}}_{\rm{s}}}{\kern 1pt} {\kern 1pt} {\kern 1pt} {{\bf{U}}_{\rm{n}}}} \right]^H},
{{\bf{R}}_{{\bf{Y}}\left( t \right)}}{\rm{ = }}\left[ {\begin{array}{*{20}{c}}
	{{{\bf{U}}_{\rm{s}}}}&{{{\bf{U}}_{\rm{n}}}}
	\end{array}} \right]{\rm{diag}}\left( {{\Sigma _{\rm{s}}},{{\bf{\Sigma }}_{\rm{n}}}} \right){\left[ {\begin{array}{*{20}{c}}
		{{{\bf{U}}_{\rm{s}}}}&{{{\bf{U}}_{\rm{n}}}}
		\end{array}} \right]^H},
\end{align}
where the eigenvalues of ${{\bf{R}}_{{\bf{Y}}\left( t \right)}}$ are sorted in 
decreasing order, ${{\bf{U}}_{\rm{s}}}{\kern 1pt}  \in {{\mathbb C}^{{N_{\rm{s}}} \times 1}}$, and ${{\bf{U}}_{\rm{n}}} \in {{\mathbb C}^{{N_{\rm{s}}} \times \left( {{N_{\rm{s}}} - 1} \right)}}$. Following \cite{cheney2001linear},  the   MUSIC  method  is applied:
 \begin{align}
\hat \mu _{{\rm{I2U}}}^{}{\rm{ = }}\mathop {\arg \max }\limits_{{\mu _{{\rm{I2U}}}}} \frac{1}{{{\bf{b}}_{\rm{s}}^H\left( {\mu _{{\rm{I2U}}}^{}} \right){{\bf{U}}_{\rm{n}}}{\bf{U}}_{\rm{n}}^H{\bf{b}}_{\rm{s}}^{}\left( {\mu _{{\rm{I2U}}}^{}} \right)}}, \label{target_direction}
 \end{align}
where the estimate of ${{\mu _{{\rm{I2U}}}}}$ can be obtained by  a simple one-dimensional search ranging from $-1$ to $1$.
\vspace{-0.3cm}
\subsection{Estimate of ToA ${{\tau _{{\rm{tot}}}}}$ and Target Parameter ${\beta _{{\rm{target}}}}$}
We collect  $N_{\rm f}$ pieces of signals in \eqref{received_signa_2} and sum them, yielding 
\begin{align}
{\bf{y}}\left( t \right)&  = {\beta _{{\rm{target}}}}{{\bf{b}}_{\rm{s}}}\left( {\mu _{{\rm{I2U}}}^{}} \right){\bf{b}}_{\rm{r}}^T\left( {\mu _{{\rm{I2U}}}^{}} \right)\sum\limits_{n = 1}^{{N_{\rm{f}}}} {{\bf{\Theta }}\left( n \right)} \notag\\
&\times{{\bf{b}}_{\rm{r}}}\left( {\mu _{{\rm{B2I}}}^{\rm{A}}} \right)s\left( {t - {\tau _{{\rm{tot}}}}} \right) + {\bf{\tilde n}}\left( t \right), \label{sum_signal_vector}
\end{align}
where ${\bf{\tilde n}}\left( t \right) \sim {\cal CN}\left( {0,{N_{\rm{f}}}{\sigma ^2}{{\bf{I}}_{{{\rm{N}}_{\rm{s}}}}}} \right)$. Then, summing all the sensor signals given in \eqref{sum_signal_vector}, i.e., 
\begin{align}
{\bar {y}}\left( t \right)& = {\rm{sum}}\left( {{\bf{y}}\left( t \right)} \right) = {\beta _{{\rm{target}}}}f\left( {\mu _{{\rm{I2U}}}^{}} \right){\bf{b}}_{\rm{r}}^T\left( {\mu _{{\rm{I2U}}}^{}} \right)\sum\limits_{n = 1}^{{N_{\rm{f}}}} {{\bf{\Theta }}\left( n \right)}\notag\\
&\times {{\bf{b}}_{\rm{r}}}\left( {\mu _{{\rm{B2I}}}^{\rm{A}}} \right)s\left( {t - {\tau _{{\rm{tot}}}}} \right) +  n\left( t \right), \label{sum_signal_scale}
\end{align} 
where $f\left( {\mu _{{\rm{I2U}}}^{}} \right){\rm{ = sum}}\left( {{{\bf{b}}_{\rm{s}}}\left( {\mu _{{\rm{I2U}}}^{}} \right)} \right)$ and $ n\left( t \right) \sim {\cal CN}\left( {0,{N_{\rm{f}}}{N_{\rm{s}}}{\sigma ^2}} \right)$. 
Here, we estimate ${{\tau _{{\rm{tot}}}}}$ and ${\beta _{{\rm{target}}}}$  by 
using DFT technique. Specifically, 
denote by $T_{\rm s}$ and ${\hat N}_{\rm s}$ the sampling  period  and the number of sampling points, respectively, and 
define  $Y\left( k \right)$, $S\left( k \right)$, and $N\left( k \right)$, $k =  - {{\hat N}_{\rm s} \mathord{\left/
		{\vphantom {N {2, \ldots ,}}} \right.
		\kern-\nulldelimiterspace} {2, \ldots ,}}{{\hat N}_{\rm s} \mathord{\left/
		{\vphantom {N {2 + 1}}} \right.
		\kern-\nulldelimiterspace} {2 + 1}}$, 
as the DFTs of $y\left( {l{T_{\rm{s}}}} \right)$, $s\left( {l{T_{\rm{s}}}} \right)$, and $n\left( {l{T_{\rm{s}}}} \right)$, $l = 0, \ldots ,{{\hat N}_{\rm{s}}-1}$, respectively. Then, the following equality holds:
\begin{align}
Y\left( k \right) = {{\bar \beta }_{{\rm{target}}}}S\left( k \right){e^{jw_{\rm tot}k}} + N\left( k \right), \label{DFT}
\end{align}
where ${w_{{\rm{tot}}}} = {{ - 2\pi {\tau _{{\rm{tot}}}}} \mathord{\left/
		{\vphantom {{ - 2\pi {\tau _{{\rm{tot}}}}} {\left( {{T_{\rm{s}}}{{\hat N}_{\rm{s}}}} \right)}}} \right.
		\kern-\nulldelimiterspace} {\left( {{T_{\rm{s}}}{{\hat N}_{\rm{s}}}} \right)}}$ and ${{\bar \beta }_{{\rm{target}}}} = {\beta _{{\rm{target}}}}f\left( {\mu _{{\rm{I2U}}}^{}} \right){\bf{b}}_{\rm{r}}^T\left( {\mu _{{\rm{I2U}}}^{}} \right)\sum\limits_{n = 1}^{{N_{\rm{f}}}} {{\bf{\Theta }}\left( n \right)} {{\bf{b}}_{\rm{r}}}\left( {\mu _{{\rm{B2I}}}^{\rm{A}}} \right)$. 
The ML  estimator for  ${w_{{\rm{tot}}}}$ and ${{\bar \beta }_{{\rm{target}}}}$ based on \eqref{DFT} is given by 
\begin{align}
&\left( {{{\hat w}_{{\rm{tot}}}},{{\hat \beta }_{{\rm{target}}}}} \right)_{\rm ML} = \notag\\
& \mathop {\arg \min }\limits_{{w_{{\rm{tot}}}},{{\bar \beta }_{{\rm{target}}}}} \sum\nolimits_{k = {{ - N} \mathord{\left/
			{\vphantom {{ - N} 2}} \right.
			\kern-\nulldelimiterspace} 2}}^{{N \mathord{\left/
			{\vphantom {N 2}} \right.
			\kern-\nulldelimiterspace} 2} + 1} {{{\left| {Y\left( k \right) - {{\bar \beta }_{{\rm{target}}}}S\left( k \right){e^{j{w_{{\rm{tot}}}}k}}} \right|}^2}}. \label{ML}
\end{align}
Taking the first-order derivative of \eqref{ML} with respect to ${{{\bar \beta }_{{\rm{target}}}}}$ and setting it to zero, we have 
\begin{align}
{{\hat \beta }_{{\rm{target}}}} = \frac{{\sum\nolimits_{k = {{ - N} \mathord{\left/
					{\vphantom {{ - N} 2}} \right.
					\kern-\nulldelimiterspace} 2}}^{{N \mathord{\left/
					{\vphantom {N 2}} \right.
					\kern-\nulldelimiterspace} 2} + 1} {{S^*}\left( k \right)Y\left( k \right){e^{ - j{w_{{\rm{tot}}}}k}}} }}{{\sum\nolimits_{k = {{ - N} \mathord{\left/
					{\vphantom {{ - N} 2}} \right.
					\kern-\nulldelimiterspace} 2}}^{{N \mathord{\left/
					{\vphantom {N 2}} \right.
					\kern-\nulldelimiterspace} 2} + 1} {{{\left| {S\left( k \right)} \right|}^2}} }}.\label{estimate_beta}
\end{align}
Then, substituting \eqref{estimate_beta} into \eqref{ML}, the ML estimator of  ${{\hat w}_{{\rm{tot}}}}$  is given by  
\begin{align}
{{\hat w}_{{\rm{tot}}}} = \mathop {\arg \min }\nolimits_{{w_{{\rm{tot}}}}} {\left| {\sum\nolimits_{k = {{ - N} \mathord{\left/
					{\vphantom {{ - N} 2}} \right.
					\kern-\nulldelimiterspace} 2}}^{{N \mathord{\left/
					{\vphantom {N 2}} \right.
					\kern-\nulldelimiterspace} 2} + 1} {{S^*}\left( k \right)Y\left( k \right){e^{ - j{w_{{\rm{tot}}}}k}}} } \right|^2}, \label{target_dealy}
\end{align}
which can be solved by  a one-dimensional search.
%\begin{algorithm}[!t]
%	\caption{Low-complexity algorithm for localizing target.}
%	\label{alg1}
%	\begin{algorithmic}[1]
%		\STATE   Estimate  direction $\mu_{\rm I2U}$  based on \eqref{target_direction}.
%		\STATE   Estimate time delay $\tau_{\rm tot}$ based on \eqref{target_dealy} and ${w_{{\rm{tot}}}} =  - 2\pi {\tau _{{\rm{tot}}}}/\left( {{T_{\rm{s}}}{{\hat N}_{\rm{s}}}} \right)$.		
%		\STATE Estimate	parameter ${\beta _{{\rm{target}}}}$ based on
%		${\hat \beta _{{\rm{target}}}}/f\left( {\hat \mu _{{\rm{I2U}}}^{}} \right){\bf{b}}_{\rm{r}}^T\left( {\hat \mu _{{\rm{I2U}}}^{}} \right)\sum\limits_{n = 1}^{{N_{\rm{f}}}} {{\bf{\Theta }}\left( n \right)} {{\bf{b}}_{\rm{r}}}\left( {\mu _{{\rm{B2I}}}^{\rm{A}}} \right)$.
%		 \STATE  Estimate  target location ${\bf q}_{\rm u}$ based on \eqref{cal_location}.
%	\end{algorithmic}
%\end{algorithm}

 Finally, based on the geometric relationship $\hat \mu _{{\rm{I2U}}}^{} = \frac{{{y_{\rm{I}}} - {y_{\rm{u}}}}}{{\left\| {{{\bf{q}}_{\rm{u}}} - {{\bf{q}}_{\rm{I}}}} \right\|}}$ and  ${{\hat \tau }_{{\rm{I2U}}}}{\rm{ = }}\frac{{\left\| {{{\bf{q}}_{\rm{u}}} - {{\bf{q}}_{\rm{I}}}} \right\|}}{c}$ and together with  \eqref{target_direction} and \eqref{target_dealy}, the  coordinates  of the target  can be calculated as\footnote{Note that only the target residing in the front half-space of IRS can be illuminated, which indicates that the solution for solving $x_{\rm u}$ is unique.} 
\begin{align}
\left\{ \begin{array}{l}
{{\hat x}_{\rm{u}}} = \sqrt {{c^2}\hat \tau _{{\rm{I2U}}}^2{\rm{ - }}{{\left( {{y_{\rm{I}}} - {{\hat y}_{\rm{u}}}} \right)}^2}{\rm{ - }}z_{\rm{I}}^2} {\rm{ + }}{x_{\rm{I}}},\\
{{\hat y}_{\rm{u}}} = {y_{\rm{I}}} - c\hat \mu _{{\rm{I2U}}}^{}{{\hat \tau }_{{\rm{I2U}}}}.
\end{array} \right. \label{cal_location}
\end{align}
%The overall  estimation procedure   is summarized in Algorithm~\ref{alg1}.
The total complexity for estimating target location is given by ${\cal O}\left( {N_{\rm{s}}^3 + {T_1}N_{\rm{s}}^2 + N_{\rm{s}}^{}{N_{{\rm{frame}}}} + \hat N_{\rm{s}}^2 + {T_2}N} \right)$.
\vspace{-0.3cm}
\section{Numerical Results}
In this section, we perform numerical experiments to evaluate the
localization performance assisted by the semi-passive IRS.  
The  BS, IRS, and target are located at $(0,0,0)$ m,  $(-10,50,2)$ m, and $(5,60,0)$ m, respectively. We consider the chirp signal $s(t)$  as $s\left( t \right) = {e^{j2\pi B\left( {t + {\nu }{t^2}} \right)}},0 \le t \le {1 \mathord{\left/
		{\vphantom {1 B}} \right.
		\kern-\nulldelimiterspace} B}$, where $B$ and $\nu $ represent the signal bandwidth and  frequency rate, respectively. 
Unless otherwise specified, other system parameters are set as follows: $N_{\rm BS}=6$, $N_{\rm s}=6$, $N_{\rm r}=50$, $N_{\rm f}=6$, ${\hat N}_{\rm s}=64$, ${{\kappa _{{\rm{RCS}}}}}=7~{\rm dBsm}$, $\nu  = {10^6}$, $B=1.5~{\rm MHz}$ , $\nu=10^6 $, ${T_{\rm{s}}}=1/\left( {B{{\hat N}_{\rm{s}}}} \right)$, ${n}_0=-150~{\rm dBm}/{\rm Hz}$, and $P_{\rm BS}=40~{\rm dBm}$. The  estimation  performance of  direction $\mu_{\rm I2U}$ and  location ${\bf q}_{\rm u}$ is evaluated by the root MSE (RMSE) \cite{shao2022target}. 
% \begin{figure*}[!t]
%	\centering
%	\begin{minipage}[t]{0.32\textwidth}
%		\centering
%		\includegraphics[width=2.5in]{powervsdirection.eps}
%		\caption{RMSE of $\mu_{\rm I2U}$ versus  $P_{\rm BS}$.}\label{fig3} 
%	\end{minipage}
%%			\hspace{3pt}
%%	\begin{minipage}[t]{0.23\textwidth}
%%		\centering
%%		\includegraphics[width=1.9in]{vsPower.eps}
%%		\caption{RMSE of ${\bf q}_{\rm u}$ versus  $P_{\rm BS}$.}\label{fig4}
%%	\end{minipage}
%			\hspace{3pt}
%	\begin{minipage}[t]{0.32\textwidth}
%		\centering
%		\includegraphics[width=2.5in]{vsreflectingelement.eps}
%		\caption{RMSE of ${\bf q}_{\rm u}$  versus   $N_{\rm r}$.}\label{fig5}
%	\end{minipage} 
%\hspace{3pt} 
%	\begin{minipage}[t]{0.32\textwidth}
%	\centering
%	\includegraphics[width=2.5in]{vsSensor.eps}
%	\caption{ RMSE of ${\bf q}_{\rm u}$  versus   $N_{\rm s}$.}\label{fig6}
%\end{minipage}
%	\vspace{-0.4cm}
%\end{figure*}

\begin{figure}[!t]
	\centerline{\includegraphics[width=2.6in]{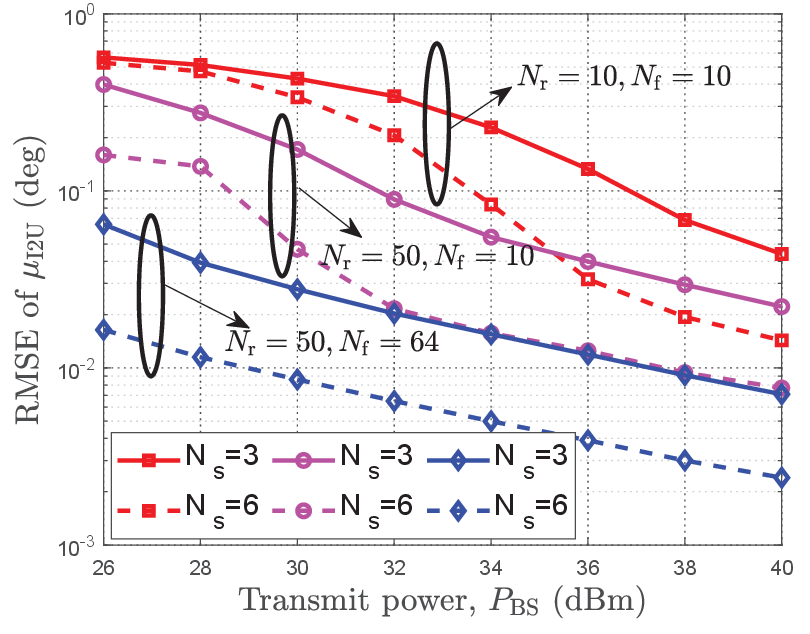}}
	\caption{RMSE of $\mu_{\rm I2U}$ versus  $P_{\rm BS}$.} \label{fig3}
		\vspace{-0.4cm}
\end{figure}
%\begin{figure}[htbp]
%	\centerline{\includegraphics[width=3in]{vsPower.eps}}
%	\caption{RMSE of ${\bf q}_{\rm u}$ versus  $P_{\rm BS}$.} \label{fig4}
	%	\vspace{-0.4cm}
%\end{figure}
In Fig.~\ref{fig3},  we evaluate the  effectiveness of   the proposed semi-passive IRS scheme by plotting the  $\mu_{\rm I2U}$-RMSE  versus  $P_{\rm BS}$  under different  $N_{\rm s}$, $N_{\rm r}$, and $N_{\rm f}$.
It is observed that   a larger number of sensors $N_{\rm s}$, a lower $\mu_{\rm I2U}$-RMSE is achieved   since  the higher  array gain can be provided at the sensors. In addition, one can observe that with a larger number of reflecting elements $N_{\rm r}$, a lower $\mu_{\rm I2U}$-RMSE is achieved. This is because a higher passive beamforming  gain can be provided by the IRS. Moreover,  the $\mu_{\rm I2U}$-RMSE can be significantly reduced with more time used for  adjusting   the IRS phase-shift vector, i.e., $N_{\rm f}$. This attributes to two reasons. First, with a higher $N_{\rm f}$, more diversity gain can be achieved since different channels are generated. Second,   with a higher $N_{\rm f}$,  a fine-grained scanning for a target can be achieved by applying the DFT-based IRS phase shift matrix, which indicates that a higher probability of illuminating the target can be obtained. 

\begin{figure}[!t]
	\centerline{\includegraphics[width=2.6in]{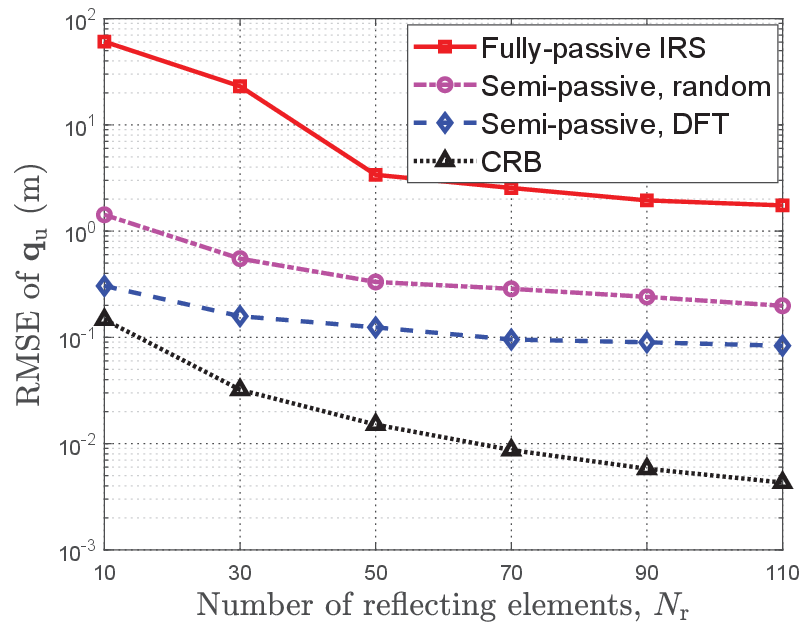}}
	\caption{RMSE of ${\bf q}_{\rm u}$  versus   $N_{\rm r}$.} \label{fig5}
	\vspace{-0.4cm}
\end{figure}
To show the superiority of the semi-passive IRS  scheme, we compare the following schemes: \textbf{1) Semi-passive, DFT}: the proposed scheme; 
\textbf{2) Semi-passive, random}: the semi-passive IRS architecture is adopted  while  the IRS phase shifts follow a  uniform distribution over $\left[ {0,2\pi } \right)$ during each frame; \textbf{3) Fully-passive, IRS}: there are  no sensors at the IRS, and  the BS  estimates target location via echo signals reflected by the IRS (For a fair comparison, the number of BS receive antennas is set the same as the number of sensors).
 \textbf{4) CRB}: the RMSE is derived based on  the  CRB given in   \eqref{CRB_location}. 
In Fig.~\ref{fig5}, we study the impact of $N_{\rm r}$ on the target location estimation  under $P_{\rm BS}=34$ dBm. It is observed that our proposed scheme outperforms the fully-passive IRS scheme. The reason is that the  signal traveling distance by the proposed scheme is much shorter than that by the fully-passive IRS scheme.  In addition, the sensing performance using the DFT-based  IRS phase-shift matrix  outperforms  that using the random-based  IRS phase-shift matrix, which shows its superiority. Furthermore, the performance gap between the proposed scheme and the CRB scheme becomes larger when $N_{\rm r}$ increases. This is because that the passive gain brought by the IRS is difficult to fully reap at the sensors since the target direction is unknown.

\begin{figure}[!t]
	\centerline{\includegraphics[width=2.6in]{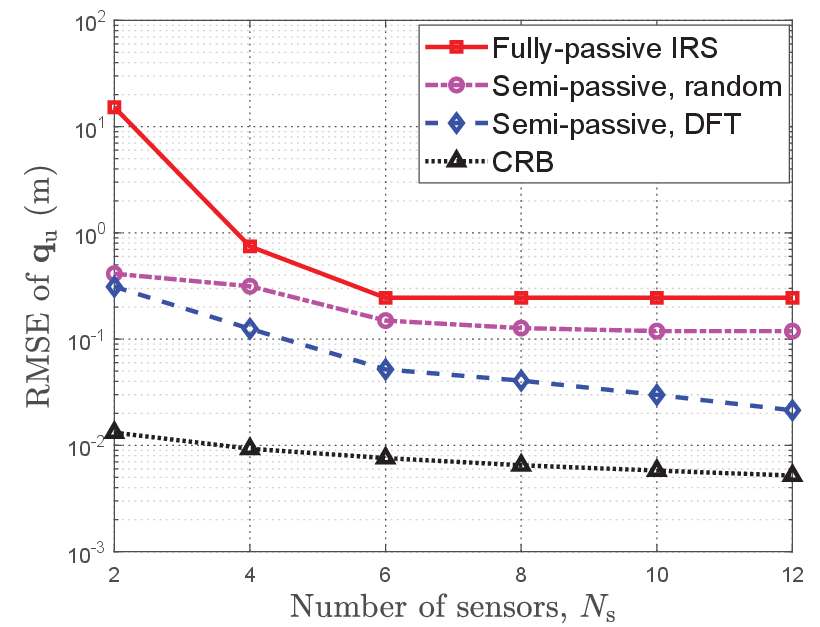}}
	\caption{RMSE of ${\bf q}_{\rm u}$  versus   $N_{\rm s}$.} \label{fig6}
	\vspace{-0.4cm}
\end{figure}

In Fig.~\ref{fig6}, the impact of $N_{\rm s}$ on the target location estimation is studied. 
%It can be seen that  the RMSE of the  proposed scheme monotonically  decreases as $N_{\rm s}$ increases due to the array gain brought by the sensors. Moreover, 
It can be seenthat the  proposed scheme outperforms the other benchmark schemes except for the CRB, especially  the performance gap becomes more pronounced when $N_{\rm s}$ is large,  which again demonstrates the superiority of the proposed scheme. In addition, the ${\bf q}_{\rm u}$-RMSE of the fully-passive IRS scheme nearly remains unchanged  while that of the proposed scheme further decreases  when $N_{\rm s} \ge 6$. This is because that the spatial direction gain can not be exploited  by the fully fully-passive IRS scheme.
Furthermore, one can observe that although there is a large performance  gap  between the proposed scheme and the CRB  when $N_{\rm s} \le 6$, the achieved ${\bf q}_{\rm u}$-RMSE of the proposed scheme approaches the CRB when $N_{\rm s}$ becomes  large. 

%for the proposed scheme, the RMSE of ${\bf q}_{\rm u}$ first rapidly decreases with $N_{\rm r}$ and then more slowly when $N_{\rm r}$ exceeds $70$.  
%
%the proposed scheme  

\section{Conclusion}
In this paper, we studied the  target localization under the  semi-passive IRS architecture.  The system parameters including the number of frames,   number of IRS reflecting elements, and  number of sensors, on  the location estimation performance based on the CRB  are theoretically analyzed. Then, we further proposed a low-complexity algorithm to estimate the location. More specifically, we first applied the MUSIC method to estimate the target direction, and  transformed the time-domain signal sequence  into the frequency-domain   based on the DFT,  then  adopted the ML criterion to jointly estimate the time delay and the  target-related coefficient. Simulation results demonstrated the superiority of  semi-passive IRS architecture  over the fully-passive IRS architecture  and showed that sub-meter level positioning accuracy can be achieved  over a long localization  range. 
%\section*{Appendix A: \textsc{derivation of ${{\bf{F}}_{{\rm{channel}}}}\left( n \right)$ in \eqref{Fisher_frame}}}\label{appendix0}
\bibliographystyle{IEEEtran}
\bibliography{IRSforlocalization}
\end{document}